# Improving Automated Hemorrhage Detection in Sparse-view CT via U-Net -based Artifact Reduction

Original Research

Johannes Thalhammer, MSc • Manuel Schultheiß, MSc • Tina Dorosti, MSc • Tobias Lasser, PhD • Franz Pfeiffer, PhD • Daniela Pfeiffer, MD • Florian Schaff, PhD

**Summary statement:** Post-processing of sparse-view cranial CT scans with a U-Net-based model allowed reduction in the number of views from 4096 to 256 with minimal impact on automated hemorrhage detection performance.

**Key points:**

1. Reducing artifacts in sparse-view cranial CTs improved automated hemorrhage detection across all investigated subtypes evidenced by the area under the receiver operator characteristic curve (AUC) values for reconstructions using 512 or fewer views (P<0.001).

2. Comparison of a U-Net and total variation-based approach for artifact reduction in sparse-view cranial CTs showed that the convolutional neural network-based approach achieved superior performance regarding artifact reduction, image quality parameters, and subsequent automated hemorrhage detection in terms of structural similarity index and AUC values across all investigated subtypes for reconstructions using 512 or fewer views (P<0.001).

## Abstract

**Purpose:** To explore the potential benefits of deep learning-based artifact reduction in sparse-view cranial CT scans and its impact on automated hemorrhage detection.

**Materials and Methods:** In this retrospective study, a U-Net was trained for artifact reduction on simulated sparse-view cranial CT scans from 3000 patients obtained from a public dataset and reconstructed with varying sparse-view levels. Additionally, the EfficientNetB2 was trained on full-view CT data from 17,545 patients for automated hemorrhage detection. Detection performance was evaluated using the area under the receiver operator characteristic curve (AUC), with differences assessed using the DeLong test, along with confusion matrices. A total variation (TV) postprocessing approach, commonly applied to sparse-view, served as the basis for comparison. A Bonferroni-corrected significance level of .001/6 = .00017 was used to accommodate for multiple hypotheses testing.

**Results:** Images with U-Net postprocessing were better than unprocessed and TV-processed images with respect to image quality and automated hemorrhage detection. With U-Net post-processing, the number of views could be reduced from 4096 (AUC: 0.97; 95% CI: 0.97-0.98) to 512 (0.97; 0.97-0.98; P<0.00017) and to 256 views (0.97; 0.96-0.97; P<.00017) with minimal decrease in hemorrhage detection performance. This was accompanied by mean structural similarity index measure increases of 0.0210 (95% CI: 0.0210-0.0211) and 0.0560 (95% CI: 0.0559-0.0560) relative to unprocessed images.

**Conclusion:** U-Net based artifact reduction substantially enhances automated hemorrhage detection in sparse-view cranial CTs.

# 1 Introduction

Intracranial hemorrhage is a potentially life-threatening disease with a 30-day fatality rate of 40.4%, accounting for 10% to 30% of annual strokes [1][2]. Prompt and accurate diagnosis is vital for optimal treatment, typically achieved through cranial CT scans [3]. However, the increased use of medical CT scans, including cranial scans, raises concerns about radiation-related health risks [4] [5]. A routine head CT scan exposes patients to a median effective dose of 2 mSv, similar to natural radiation accumulated per year. This affects not only the brain but also nearby areas [5] [6].

Besides lowering the X-ray tube current, sparse-view CT is a promising approach to reduce dose by decreasing the number of views. However, this reduction causes artifacts in the filtered back projection (FBP) reconstructed images. To restore image quality for accurate diagnosis, suitable processing methods are needed. In the past, compressed sensing (CS) and iterative reconstruction approaches have been widely investigated, which generally minimize a CS-based regularization term as well as a data-fidelity term to ensure data consistency [7] [8] [9]. These approaches have been proven to yield good results in terms of reducing image noise and artifacts. However, they are also computationally demanding due to the repeated update steps during iterative optimization. Additionally, these methods require parameter optimization and can alter image texture.

Recently, machine learning approaches using deep neural networks have gained substantial attention. In the context of artifact reduction in sparse-view CT, it has been shown that convolutional neural networks (CNNs) are able to achieve excellent results with comparably low computational effort during inference [10] [11] [12]. Approaches that combine deep learning and iterative techniques for sparse-view artifact reduction can also be found in the literature [13] [14].

In parallel, extensive research has been conducted on the application of deep learning techniques for automated detection and classification of pathological features in CT images, and many systems are already Food and Drug Administration-approved and CE-certified [15] [16] [17]. Typically, these algorithms are trained on standard-dose data, which restricts their applicability to dose-reduced or sparse-view data. This is because the additional noise and artifacts in sparse-view data are expected to negatively impact the reliability of these systems.

Therefore, the aim of the current study is twofold. Firstly, we aim to explore the potential benefits of deep learning-based artifact reduction in sparse-view cranial CTs. Secondly, we assess whether this approach can enhance the performance of automated hemorrhage detection.

## 2 Materials and Methods

This retrospective study was exempt from institutional review board review due to the use of publicly available data. The code is available at: https://github.com/J-3TO/Sparse-View-Cranial-CT-Reconstruction.

### 2.1 Network Architectures

Given the parallel beam geometry of our dataset, where streak artifacts are expected to be contained within the individual 2D slices without extending into neighboring ones, we opted for a 2D U-Net architecture in our experiments, considering it the most efficient option. Figure 1 illustrates the U-Net architecture for artifact reduction for a 512x512 input [18]. The initial input is downsampled by encoder blocks consisting of convolutional layers and strided convolutions, decreasing the spatial resolution down to the size of the bottleneck feature maps. The subsequent upsampling is performed through encoder blocks, consisting of, again, convolutional layers and transposed strided convolutions, which were adapted from Guo et al. [19]. Skip connections connect the encoder and decoder parts. The final network output is generated by summing the initial input image with the output of the last decoder block. In depth details about the architecture can be found in the legend of Figure 1. The model was initialized with random weights. For hemorrhage detection, the EfficientNetB2 was used, initialized with ImageNet pretrained weights [20]. Both models were implemented in Keras (version 2.0.4) [21].

## 2.2 Datasets

This study used the RSNA 2019 Brain CT Hemorrhage Challenge dataset, with each CT image annotated for hemorrhage presence and subtype (subarachnoid, intraventricular, subdural, epidural, and intraparenchymal) [22] [23]. The dataset consists of CT scans from 18,938 different patients. After filtering out images that either did not have a resolution of 512x512 pixels, or led to various errors during data generation, the dataset was narrowed down to 18,545 patients. The list of patients was randomly shuffled via the Fisher-Yates algorithm, implemented in Python's built-in random module (Python 3.8.10) [24]. The first 1,000 patients were then selected for testing and excluded from all further training.

For U-Net training, the next 3,000 patients were selected from the patient list. Prior experiments have indicated that this number is sufficient to effectively train the U-Net without experiencing overfitting problems [10][12]. The first 2,400 patients (80%) served as training data, the remaining 600 patients (20%) as validation data. The CT pixel values of the dataset were encoded as either 12-bit or 16-bit unsigned integers. To create sparse-view CTs, we clipped the 16-bit images to the potential 12-bit image range of [0-4,095] and divided all images by 4,095 to normalize them to range [0-1]. Sinograms with 4,096 views were created under parallel beam geometry from the CT dataset using the Astra Toolbox (version 2.1.0) [25]. Reference standard images were reconstructed from 4,096 views using FBP. Six sparse-view subsets were generated using FBP with 64, 128, 256, 512, 1,024 and 2,048 views, respectively.

Due to overfitting concerns, the remainder of the dataset, except the test split, was used for training the EfficientNetB2 (17,545 patients). Individual images were scaled to Hounsfield Units (HUs) and clipped to the diagnostically relevant brain window [0-80] HU. To meet the requirements of the pre-trained ImageNet, images were rescaled to [0-255], resized to 260x260 pixels by bilinear resizing and transformed to three channel images by concatenating three neighboring CT images [26].

Figure 2 depicts the data selection process and the distribution of the labelled hemorrhage subtypes. For external testing, the CQ500 dataset (mean age, 48.13 (range, 7-95), 36.25% (178) female, 63.75% (313) male) derived from the Centre for Advanced Research in Imaging, Neurosciences and Genomics (CARING) in New Delhi, India, was used [27]. The file list was shuffled using the Fisher-Yates algorithm, and the first 10,000 images were selected. These images were pre-processed in the same manner as the RSNA dataset.

## 2.3 Training

The U-Net was trained separately for each sparse-view subset, using sparse-view images as input and full-view images as reference standard. It was trained with mean squared error loss for 75 epochs with a mini-batch size of 32. Randomly selected 256x256 patches from the images were rotated by 0°, 90°, 180°, or 270°. The learning rate was $lr=10^{-4}/(epoch+1)$. Hyperparameters were selected based on a trial-and-error approach, ensuring that the loss curves exhibited fast convergence. The EfficientNetB2 was trained with 5-fold cross-validation and binary cross-entropy loss with a mini-batch size of 32 for 15 epochs. The learning rate followed a cosine annealing schedule with warm restarts after epoch one, three, and seven with an initial learning rate of $5 \cdot 10^{-4}$ and minimal learning rate of $10^{-5}$. The model with the lowest validation loss was selected for each split. The final predictions for each image were obtained by calculating the arithmetic mean of each class from the outputs of the five different splits.

## 2.4 Total Variation

In this work, we used the isotropic total variation (TV) method by Chambolle for artifact reduction (scikit-image version 0.19.3) [28] [29]. The optimal weight for each sparse-view subset was determined by randomly sampling 1,000 images from the U-Net training set and iterating through weights ranging from 0.001 to 1.000 in 0.001 increments. We identified a global maximum within this weight range for each subset, suggesting the range was reasonable. The weights that yielded the best score for the structural similarity index measure (SSIM) were selected to calculate the metrics on the test set [30]. We also explored using the peak-signal-to-noise ratio (PSNR) as an optimization metric. The resulting images however, exhibited substantial visual degradation compared with SSIM optimization.

## 2.5 Saliency Maps

The saliency maps were obtained by computing the gradient of the class score with respect to the input image, as described in [31]. To quantify the saliency maps, we shuffled (Fisher-Yates) the test set images containing intraparenchymal hemorrhages and selected the first 100. The hemorrhage regions were manually segmented and verified by a board-certified radiologist, D.P., who has 16 years of experience. For the selected images, we then calculated the saliency map for each sparse-view level with and without U-Net post-processing, respectively. The segmented masks were used to compute the ratio of the sum of saliency map values within the mask to the sum of values outside the mask.

## 2.6 Statistical Analysis

Image quality of the different post-processing methods was compared using SSIMs, PSNRs, and (signal-to-noise ratios) SNRs (scikit-image version 0.19.3) [29]. To ensure a focus on diagnostically relevant pixels while excluding potentially distorting background pixels, both metrics were applied within a mask of the intracranial region. This mask was generated using the CNN developed and trained by Cai et al. [32]. A Shapiro-Wilk test showed that the metrics did not follow a normal distribution. Subsequently, 95% confidence intervals of the mean were calculated using bootstrapping with 1,000 resamples. The Wilcoxon signed-rank test was performed to examine significant differences between no post-processing and TV or U-Net post-processing. A Bonferroni corrected significance level to account for the multiple hypothesis tests of 0.00033 (0.001/3) was used for the p-values [33].

To assess inference speed, 100 images from each sparse-view subset were selected from the test set, and the individual inference times for TV and U-Net post-processing were measured on an NVIDIA GeForce RTX 3090 with 24 GB VRAM. This process was repeated 50 times, and 95% confidence intervals were calculated using bootstrapping with 1,000 resamples.

To quantify hemorrhage detection performance, empirical area under the receiver operator characteristic curve (AUC) values, including 95% CIs, were estimated as described by DeLong et al [34]. Thereby, the algorithmic implementation by Sun and Xu adapted to Python 3.8.10 was used [35]. Statistical differences in AUCs between sparse-view and full-view datasets, as well as between different post-processing methods, were evaluated using the DeLong two-sided test. Again, Bonferroni correction was applied, leading to a significance level of p=0.00017 (0.001/6) for comparing the six sparse-view subsets with the full-view data and a significance level of p=0.00033 (0.001/3) for comparing raw FBP, TV post-processing, and U-Net post-processing. Confusion matrices were generated with Scikit-learn (version 1.1.3); the discrimination thresholds were selected for each subset to maximize the geometric mean of the true positive and true negative rates [36]. The saliency map ratios with and without U-Net post-processing were compared in the same manner as the SSIM and PSNR values by using the Wilcoxon signed-rank test and calculating 95% confidence intervals of the mean by bootstrapping with 1,000 resamples. If not stated otherwise, Scipy version 1.4.1 was used for the statistical analysis [37].

# 3 Results

## 3.1 Artifact Reduction

Figure 3a-e shows a CT image from the test set, reconstructed with varying number of views. The full-view (4096 views) image showed a clearly visible intraparenchymal hemorrhage. The intraventricular subtype was also discernible, although it was more challenging to detect. The image reconstructed from 512 views showed decreased image quality and presence of artifacts appear, but the hemorrhages could still be identified. In the image reconstructed from 256 views, streak artifacts become pronounced, making it challenging to distinguish small features. Although the intraparenchymal subtype was still discernible, the intraventricular subtype was barely recognizable. Images reconstructed from 128 and 64 views show severe streak artifacts and distortion of the brain tissue. The hemorrhages could not be identified. Figure 3f-i shows the U-Net predictions of the sparse-view images, revealing a clear reduction in streak artifacts compared with the respective input images (Figure 3b-e). With increasingly sparse-sampled input, the prediction also tended to become smoother (i.e., sharp image features were not retained). However, image quality of the sparse-view CTs still improved, and the similarity between predictions and full-view images increased. The contours of the hemorrhage can be recognized until the 256-view and 128-view predictions for intraparenchymal and intraventricular subtypes, respectively.

## 3.2 Comparison with Total Variation (TV)

For comparison, we also implemented TV-based artifact reduction, which is commonly used to address undersampling [38] [39]. Figure 4a-f displays results of an image labeled "healthy" from the test set, reconstructed from a varying number of views. In the 2048-view reconstruction no artifacts were visible. When the number of views was further reduced, the image quality deteriorated with only the skull shape discriminable in the 64-view reconstruction. Figure 4g-m shows the respective U-Net predictions of the sparse-view CT images. Consistent with figure 3, the U-Net reduced the artifacts considerably. Again, the resulting images were increasingly smoothed as the number of views decreased. In Figure 4n-s the results of TV post-processing are depicted, which could reduce the artifacts well down to 256 views sparse sampling. However, the results for 128- and 64-sparse view data were inferior compared to the results from U-Net post-processing. An additional example is presented in Figure 5, showing images labeled with a subarachnoid hemorrhage.

Table 1 presents the mean SSIM, PSNR, and SNR values of the reconstructed images calculated on the RSNA test set and the CQ500 dataset, with individual values calculated in reference to the 4096-view images. Both, U-Net and the TV post-processing quantitatively increased the SSIM by reducing streak artifacts compared with raw FBP reconstructions. This aligns with visual results in figures 3, 4, and 5. U-Net post processing also enhanced PSNR values, whereas TV-processing resulted in decreased PSNR values for specific subsets, specifically down to and including the 512-view subset for the RSNA test split and the 2048-view subset of the CQ500 dataset, when compared with the respective raw FBP images. The PSNR values of the remaining subsets improved after TV-processing. Direct comparison revealed U-Net's stronger performance in all cases, with statistically significant differences between post-processing for each subset ($P<.00033$).

The mean inference speed of the TV method and the U-Net were also compared for 100 images per sparse-view subset. For the 64-, 128-, 256-, 512-, 1024-, and 2048-view data, the U-Net took 2.254 (95% CI: 2.249-2.258), 2.244 (2.24-2.249), 2.209 (2.206-2.214), 2.206 (2.203-2.211), 2.213 (2.205-2.244) seconds, and the TV algorithm took 46.886 (95% CI: 46.595-47.19), 30.84 (30.704-30.99), 17.091 (17.02-17.16), 7.393 (7.343-7.449), 7.237 (7.196-7.275), 5.096 (5.069-5.129) seconds, respectively.

### 3.3 Detection of Hemorrhage Subtypes

Finally, we evaluated the impact of artifact reduction on automated hemorrhage detection by examining the outcomes of the EfficientNetB2 hemorrhage detection network. Figure 6 demonstrates the AUC values for the raw images (blue) and the images post-processed by either TV (green) or the U-Net (orange) for varying levels of subsampling for each subtype. If not stated otherwise, the given AUC values in the following text refer to the predicted "any" class, stating if a hemorrhage was present in the image, regardless of subtype. Overall, detection performance decreased as the number of views used for the reconstruction decreased. When reducing the number of views from 4096 (AUC=0.97; 95% CI: 0.97-0.98) to 1024 (AUC=0.96; 95% CI: 0.96-0.97), there was a slight decrease in performance (P=0.01 for epidural, P<.00017 for other subtypes), which further decreased with fewer views. In TV post-processed images, the AUC values slightly decreased until 512-views (AUC=0.96; 95% CI: 0.96-0.96; P=0.00033 for epidural, P<.00017 for other subtypes) and decreased significantly for fewer views (256-views: AUC=0.91; 95% CI: 0.91-0.92, P<.00017). With the U-Net, there was a minimal decrease in detection performance as the number of views reduced from 4096 views to 512 views (AUC=0.97; 95% CI: 0.97-0.98; P=0.62 for epidural, P<.00017 for other subtypes), and to 256 views (AUC=0.97; 95% CI: 0.96-0.97; P=0.17 for epidural, P<.00017 for other subtypes). Below 256 views, a noticeable decline is visible (128-views: AUC=0.94; 95% CI: 0.94-0.95, P<.00017). For all the cases reconstructed with 1024 views or fewer, except the epidural subtype, the AUC values obtained from U-Net post-processed images surpass those of TV-processed and raw sparse-view images significantly (P<.0003). For the epidural subtype, a significant difference is only observable for 512 views or fewer (512-views: U-Net: AUC=0.88; 95% CI: 0.84-0.92; TV: AUC=0.85; 95% CI: 0.80-0.90). Notably, the detection of epidural hemorrhages is generally poorer compared with other subtypes. The individual AUC values of all subtypes, and the corresponding ROC curves are given in the supplementary material in Table S1 and Figure S1, respectively. All the individual p-values can be found in the supplementary table S2.

Table 2 shows the confusion matrices for the 'any' class of CT images reconstructed with varying numbers of views, without post-processing (FBP) and with either TV or U-Net post-processing. The decrease in detection performance and the impact of post-processing agree with the results in figure

6. The confusion matrices for all classes can be found in the supplementary materials.

Examples of individual image detection results are depicted in Figure 7. The presence of an intraparenchymal hemorrhage remained undetected in the sparse-view reconstructions, despite U-Net or TV post-processing. However, for the subarachnoid hemorrhage, TV post-processing, and for the intraventricular hemorrhage, U-Net post-processing, yielded correct detection results. For the first healthy case, both sparse-view and post-processed images were erroneously classified as hemorrhage positive. Nevertheless, for the second healthy case, TV post-processing resulted in a true negative detection, while U-Net post-processing lead to a true negative detection in the third healthy case presented.

## 3.4 Saliency Maps

Figure 8 shows the saliency maps for the images of figure 3 with regards to the 'any' class. The rectangles are positioned identically to those in figure 3, indicating the location of the hemorrhages. For images reconstructed down to and including 256-view sparse-sampling (subfigures a) - c)), the network primarily focused on the area of the intraparenchymal hemorrhage for its prediction. However, for reconstructions with fewer views (subfigures d) and f)), such a focused area was no longer discernible. Conversely, for the U-Net post-processed images (subfigures f) - i), all the saliency maps focused on the subarachnoid hemorrhage area.

The results of the quantitative saliency map analysis are displayed in Table 3. When using 128 views or fewer, a significant difference was discernible between raw FBP reconstructions and U-Net post-processed ones (P<0.001). This agrees with the results shown in figure 6 d) and figure 8.

## 4 Discussion

In this work, we investigated CNN-based artifact reduction in sparse-view cranial CTs and the subsequent impact on automated hemorrhage detection. We demonstrated that a deep CNN, specifically a U-Net architecture, leads to substantial improvements in the visual quality of sparse-view cranial CTs. Evaluation using the SNR, PSNR and SSIM metrics quantitatively confirmed the enhanced image quality achieved by the network. Furthermore, we trained a hemorrhage detection network on full-view CT images and applied it to sparse-view images with and without post-processing by the U-Net. The results showed that the U-Net enabled a reduction in views from 4096 to 512 with minimal impact on detection performance, and to 256 views with only a slight performance decrease. Compared with the 4096-view image, this would correspond to a respective dose reduction of 87.5% (1-512/4096) and 93.75% (1-256/4096). However, even without post-processing, a meaningful decline in detection performance was observed for subsets using fewer than 2048 views. This suggests that using fewer views for the full-view image would also be sufficient, subsequently mitigating the stated dose reduction. In comparison, Prasad et al. demonstrated a 50% reduction in radiation dose for low-dose chest CTs and We et al. a 46% dose reduction for cranial CTs, while maintaining diagnostic integrity [40][41]. Additionally, we compared the results of the U-Net with an analytical approach based on TV. The U-Net had superior performance compared with TV post-processing with respect to image quality parameters, inference speed, and automated hemorrhage detection.

We selected the U-Net architecture for its multiscale encoder-decoder structure with skip-connections, allowing it to efficiently solve image-to-image problems without requiring complex training optimizations. The excellent results with U-Net based approaches in sparse-view artifact reduction, exemplified by studies such as those conducted by Han et al., Jin et al., or Genzel et al., whose method notably secured first place in the AAPM DL-sparse-view CT challenge, further motivated our adaption of this architecture [10] [11] [42]. The artifact reduction performance of our network was on par with these reported methods. For instance, the U-Net proposed by Jin et al. a U-Net with an additional skip connection from input to output and reported signal-to-noise ratio (SNR) improvements of 15.31 dB and 11.18 dB for 50 views and 143 views, respectively, compared to raw FBP reconstructions on their "biomedical dataset" [11]. We observe SNR improvements of 22.11 dB and 20.1 dB for the 64-view and 128-view images respectively compared to raw FBP reconstructions for the RSNA test split. Han et al., extended the U-Net architecture to satisfy the frame condition and reports PSNR improvements of 11.57 dB and 9.97 dB for 60 views and 120 views respectively [10]. In comparison, we observed PSNR improvements of 22.11 dB and 20.01 dB for our 64-view and 128-view reconstructions, respectively. It is essential to note that drawing conclusions about method superiority from these values is difficult given the differences in training datasets and data preprocessing methods, which can substantially influence the results. The success of these architectures, including our U-Net, can be attributed to their multi-resolution feature. The exponentially large receptive field due to the pooling and unpooling layers makes it possible to handle streak artifacts that occur in sparse-view CT and typically spread over a large portion of the image. We chose a 2D approach due to the parallel beam geometry of our data. In cases with different geometries, such as cone beam, it might be worthwhile to explore a computationally more expensive 3D U-Net variant in future works.

The U-Net demonstrated robustness by substantially improving image quality across all investigated levels of subsampling on both the RSNA dataset and the external CQ500 dataset. This was evidenced by the calculated PSNR and SSIM values. The TV approach was also able to significantly improve the SSIM values of the sparse-view data. Interestingly, no clear trend in the PSNR values could be identified for TV processing. This is most likely because the weights for the TV method were set to optimize the SSIM, rather than PSNR, as described in section 2.4. The results of automated hemorrhage detection further validated the importance of artifact reduction, with substantial improvements observed in AUC values and confusion matrices. Saliency maps provided additional support to these findings by revealing that pronounced artifacts in sparse-view images disrupted the detection process, leading to increased false positives and false negatives. By reducing these artifacts through post-processing, these effects can be mitigated and the detection performance can be maintained, without the need for specific training on sub-sampled data.

It is important to acknowledge some limitations of this study. The sparse-view data used in our study was retrospectively generated under simplified conditions from CT volumes, which may not fully capture the complexity of real-world scenarios. Additionally, the training dataset had imbalances in terms of negative cases and not all hemorrhage subtypes were represented equally. This might explain the relatively poor performance in classifying the epidural subtype compared to the other subtypes. Furthermore, analyzing the patient demographics across data splits was not possible due to the unavailability of such information in the RSNA dataset.

In summary, our findings highlight the importance of employing appropriate post-processing techniques to achieve optimal image quality and diagnostic accuracy while minimizing radiation dose. Furthermore, our study demonstrates that leveraging deep learning methods for artifact reduction can lead to significant improvements in hemorrhage detection on sparse-view cranial CTs. This has promising implications for rapid automated hemorrhage detection on sparse-view cranial CT data to assist radiologists in routine clinical practice. Subsequent studies focused on evaluating U-Net-based artifact reduction using clinically measured sparse-view data are essential for the successful integration of this approach into clinical practice.

Tables

| RSNA test split | | | | | | |
|---|---|---|---|---|---|---|
| SSIM | 2048 views | 1024 views | 512 views | 256 views | 128 views | 64 views |
| FBP | 1.000 (1.000-1.000) | 0.994 (0.994-0.994) | 0.979 (0.979-0.980) | 0.943 (0.942-0.943) | 0.815 (0.814-0.817) | 0.655 (0.652-0.657) |
| U-Net | 1.000 (1.000-1.000) | 1.000 (1.000-1.000) | 1.000 (1.000-1.000) | 0.999 (0.999-0.999) | 0.998 (0.998-0.998) | 0.997 (0.997-0.997) |
| TV | 1.000 (1.000-1.000) | 0.999 (0.999-0.999) | 0.999 (0.999-0.999) | 0.998 (0.997-0.998) | 0.991 (0.991-0.991) | 0.975 (0.974-0.975) |
| PSNR [dB] | 2048 views | 1024 views | 512 views | 256 views | 128 views | 64 views |
| FBP | 74.275 (74.191-74.345) | 63.002 (62.859-63.153) | 57.264 (57.113-57.41) | 47.252 (47.159-47.351) | 38.685 (38.618-38.765) | 32.544 (32.479-32.611) |
| U-Net | 78.099 (78.049-78.15) | 72.504 (72.432-72.574) | 68.661 (68.601-68.728) | 63.109 (63.062-63.157) | 58.783 (58.743-58.82) | 54.651 (54.605-54.691) |
| TV | 66.526 (66.487-66.561) | 63.177 (63.132-63.22) | 60.073 (60.044-60.098) | 52.322 (52.293-52.35) | 44.882 (44.852-44.909) | 38.357 (38.327-38.387) |
| SNR [dB] | 2048 views | 1024 views | 512 views | 256 views | 128 views | 64 views |
| FBP | 48.019 (47.864-48.17) | 36.801 (36.614-37.028) | 31.051 (30.858-31.264) | 20.985 (20.832-21.15) | 12.395 (12.242-12.538) | 6.249 (6.11-6.395) |
| U-Net | 51.82 (51.693-51.962) | 46.243 (46.099-46.385) | 42.386 (42.252-42.533) | 36.828 (36.704-36.961) | 32.495 (32.375-32.633) | 28.356 (28.22-28.475) |
| TV | 40.236 (40.105-40.359) | 36.897 (36.765-37.029) | 33.777 (33.652-33.9) | 26.015 (25.889-26.135) | 18.574 (18.452-18.702) | 12.054 (11.923-12.174) |
| CQ500 dataset | | | | | | |
| SSIM | 2048 views | 1024 views | 512 views | 256 views | 128 views | 64 views |
| FBP | 1.000 (1.000-1.000) | 0.987 (0.987-0.987) | 0.957 (0.956-0.957) | 0.876 (0.876-0.877) | 0.681 (0.68-0.683) | 0.479 (0.477-0.481) |
| U-Net | 1.000 (1.000-1.000) | 0.999 (0.999-0.999) | 0.997 (0.997-0.997) | 0.994 (0.994-0.994) | 0.992 (0.991-0.992) | 0.988 (0.988-0.989) |
| TV | 1.000 (1.000-1.000) | 0.998 (0.998-0.998) | 0.996 (0.995-0.996) | 0.991 (0.991-0.991) | 0.976 (0.976-0.976) | 0.948 (0.947-0.948) |
| PSNR [dB] | 2048 views | 1024 views | 512 views | 256 views | 128 views | 64 views |
| FBP | 67.941 (67.896-67.98) | 51.191 (51.116-51.269) | 45.530 (45.456-45.603) | 39.783 (39.735-39.836) | 33.457 (33.411-33.503) | 28.252 (28.205-28.296) |
| U-Net | 73.380 (73.337-73.418) | 65.267 (65.208-65.329) | 61.363 (61.303-61.429) | 56.771 (56.719-56.827) | 53.438 (53.395-53.485) | 50.406 (50.361-50.461) |
| TV | 62.284 (62.249-62.314) | 57.817 (57.779-57.851) | 56.203 (56.161-56.246) | 49.875 (49.84-49.908) | 42.644 (42.611-42.676) | 35.843 (35.818-35.871) |
| SNR [dB] | 2048 views | 1024 views | 512 views | 256 views | 128 views | 64 views |
| FBP | 39.405 (39.238-39.577) | 28.137 (27.885-28.383) | 22.388 (22.14-22.599) | 12.364 (12.184-12.568) | 3.800 (3.619-3.974) | 2.348 (2.509-2.180) |
| U-Net | 43.224 (43.052-43.385) | 37.627 (37.47-37.797) | 33.771 (33.607-33.931) | 28.217 (28.065-28.38) | 23.888 (23.739-24.039) | 19.756 (19.599-19.907) |
| TV | 31.641 (31.477-31.773) | 28.297 (28.132-28.458) | 25.178 (25.019-25.318) | 17.428 (17.282-17.588) | 9.990 (9.855-10.139) | 3.474 (3.314-3.617) |

*Table 1*: Structural similarity index measurement (SSIM) and peak signal-to-noise-ratio (PSNR) values of the region of interest of the sparse-view CT images of the RSNA test split and the CQ500 dataset.

Note.— The images were reconstructed by filtered back projection (FBP) from varying numbers of views and then not post-processed, post-processed by the U-Net, or post-processed by the total variation-based method (TV). The differences between these three methods were significant at P<0.00033 for all subsets. The data are presented as means with 95% CI in parentheses. *SSIM = Structural Similarity Index Measurement; PSNR = Peak Signal-to-Noise-Ratio; FBP = Filtered Back Projection; TV = Total Variation*

*Table 2:* Confusion matrices for hemorrhage detection performance of the full-view images and reconstructed images with varying numbers of views without post-processing (FBP) and postprocessing with either U-Net or TV.

| Nr. of views | | 4096 | | 2048 | | 1024 | | 512 | | 256 | | 128 | | 64 | |
|---|---|---|---|---|---|---|---|---|---|---|---|---|---|---|---|
| | actual predicted | yes | no | yes | no | yes | no | yes | no | yes | no | yes | no | yes | no |
| FBP | yes | 4,684 | 2,287 | 4,703 | 2,456 | 4,588 | 3,343 | 4,431 | 6,580 | 4,167 | 9,398 | 3,720 | 11,937 | 3,630 | 12,952 |
| | no | 455 | 31,999 | 436 | 31,830 | 551 | 30,943 | 708 | 27,706 | 972 | 24,888 | 1,419 | 22,349 | 1,509 | 21,334 |
| U-Net | yes | | | 4,686 | 2,293 | 4,659 | 2,104 | 4,673 | 2,388 | 4,568 | 2,541 | 4,344 | 3,277 | 4,025 | 6,287 |
| | no | | | 453 | 31,993 | 480 | 32,182 | 466 | 31,898 | 571 | 31,745 | 795 | 31,009 | 1,114 | 27,999 |
| TV | yes | | | 4,607 | 2,370 | 4,585 | 2,390 | 4,486 | 2,697 | 4,309 | 6,129 | 3,887 | 9,152 | 3,691 | 13,881 |
| | no | | | 532 | 31,916 | 554 | 31,896 | 653 | 31,589 | 830 | 28,157 | 1,252 | 25,134 | 1,448 | 20,405 |

*Note—Images were reconstructed with FBP from 2048 to 64 views.* Each 4x4 field represents an individual confusion matrix corresponding to one sparse-view dataset and one processing method. *FBP = Filtered Back Projection; TV = Total Variation*

Table 3: Ratios of the saliency map values inside the intraparenchymal mask to the values outside the mask without (FBP) and with U-Net post-processing.

|  | 2048 views | 1024 views | 512 views | 256 views | 128 views | 64 views |
|---|---|---|---|---|---|---|
| FBP | 0.094 (0.084-0.104) | 0.092 (0.082-0.104) | 0.100 (0.088-0.115) | 0.086 (0.074-0.098) | 0.052 (0.042-0.064) | 0.016 (0.013-0.019) |
| U-Net | 0.094 (0.085-0.106) | 0.094 (0.085-0.104) | 0.094 (0.084-0.104) | 0.098 (0.087-0.109) | 0.104 (0.093-0.118) | 0.107 (0.093-0.125) |
| P-value | 0.47 | 0.18 | 0.97 | 0.02 | <0.001 | <0.001 |

*Note--* The last row depicts the p-values associated with the Wilcoxon two-sided statistical test, comparing the FBP and U-Net values for statistical differences. FBP = Filtered Back Projection; TV = Total Variation

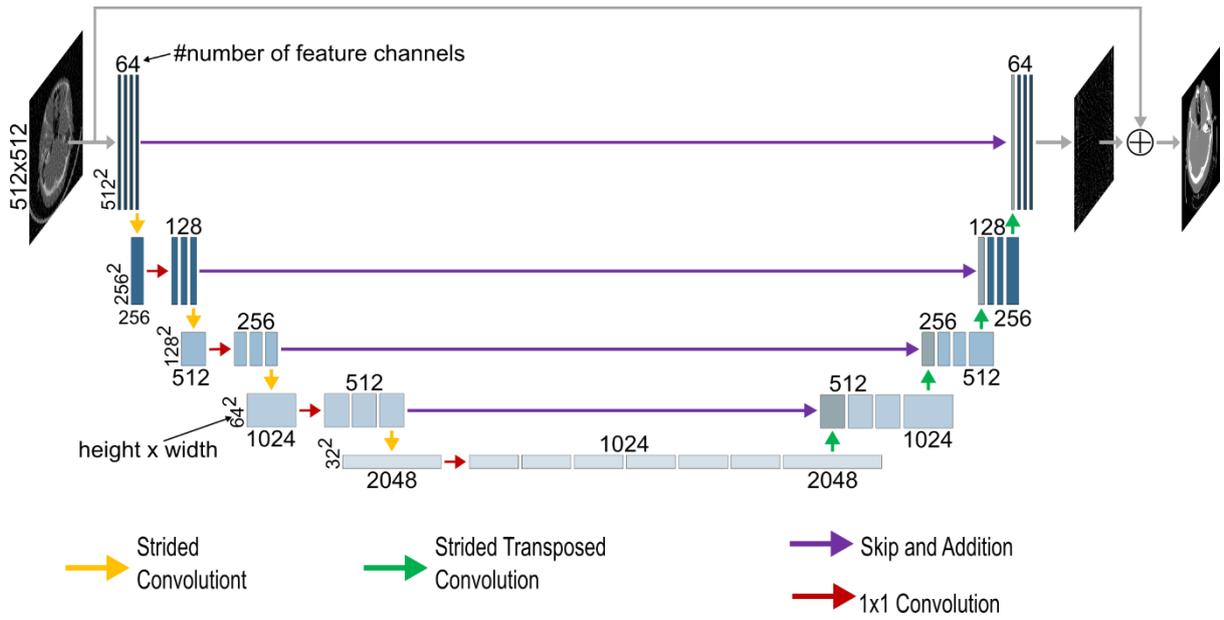

*Figure 1:* The architecture of the used U-Net for a 512x512 input. If not stated otherwise, each block of feature channels is connected to the previous one by a 3x3 convolution.

Explanatory Note: The vertical numbers represent the resolution of the feature channels at each level, while the horizontal numbers indicate the number of feature channels in each block. Initially, the input undergoes processing through a block of four convolutional layers with 64 feature channels, followed by four encoding blocks. In each encoding block, the input is downsampled using a strided convolution with stride 2x2 and a 1x1 convolution, followed by three convolutional layers with kernel size 3x3. After the four encoding blocks, a bottleneck feature map of size 32x32 with 1024 feature channels is obtained. The feature maps then undergo expansion through four decoding blocks. In each block, the input is processed by three convolutional layers with kernel size 3x3, followed by upsampling using a strided transposed convolution with stride 2x2. This upsampling approach was adapted from Guo et al. [20]. The feature maps are then added with the corresponding feature maps from the encoding path. Following the decoding blocks, three 3x3 convolutional layers with 64 output channels and one convolutional layer with one output channel are applied. The final output is obtained by adding the initial input to this feature map.

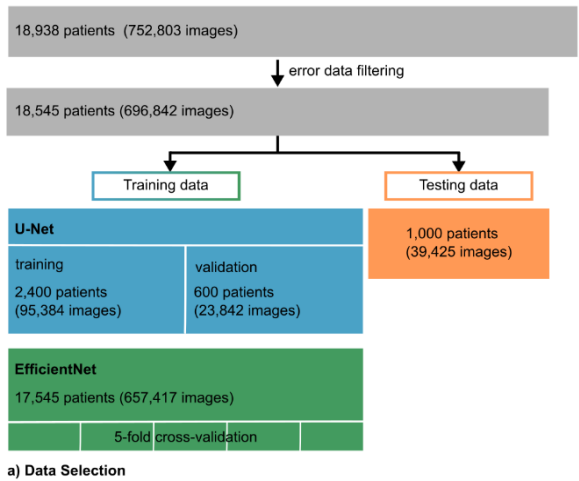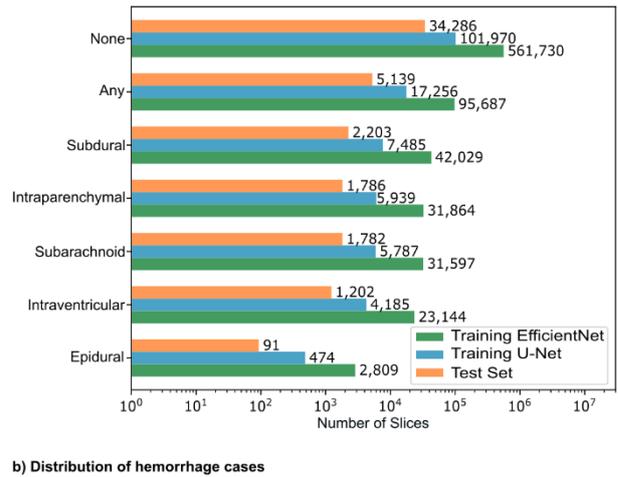

*Figure 2:* a) Flowchart of the data selection process. b) Distribution of the labeled hemorrhage subtypes in the used data split of the EfficientNetB2, the U-Net and the test set. Note the logarithmic scaling of the x-axis.

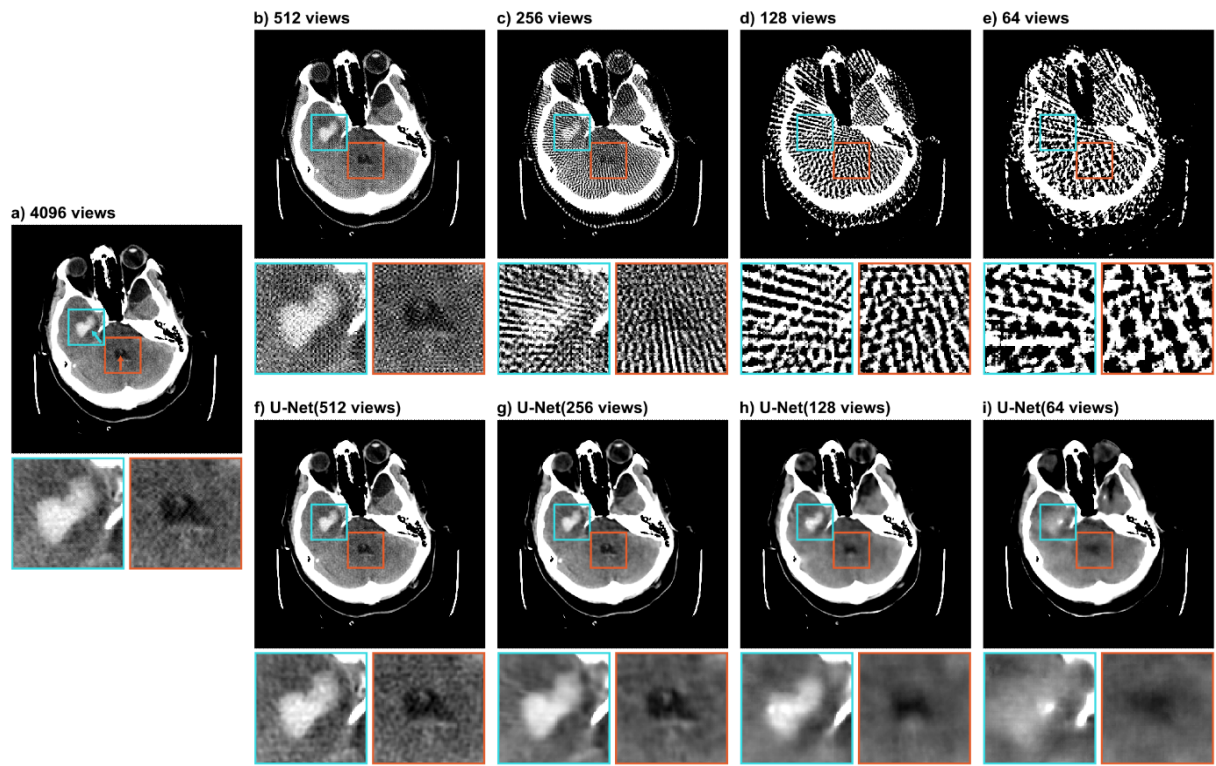

*Figure 3:* CT image (512x512 pixels) from the test set labeled with an intraparenchymal (cyan arrow) and an intraventricular (orange arrow) hemorrhage. The top row displays raw images, and the bottom row demonstrates artifact reduction by the U-Net. The labelled intraparenchymal and intraventricular hemorrhages are shown in detail in the zoomed-in extracts. Image a) shows the image reconstructed from 4096 views. Images b) - e) show the same image reconstructed from 512, 256, 128, and 64 views, respectively. Images f)- i) show the U-Net predictions of the corresponding sparse-view images in the upper row. All images are presented in the brain window ranging from 0 HU to 80 HU. Both inserts are 80x80 pixels.

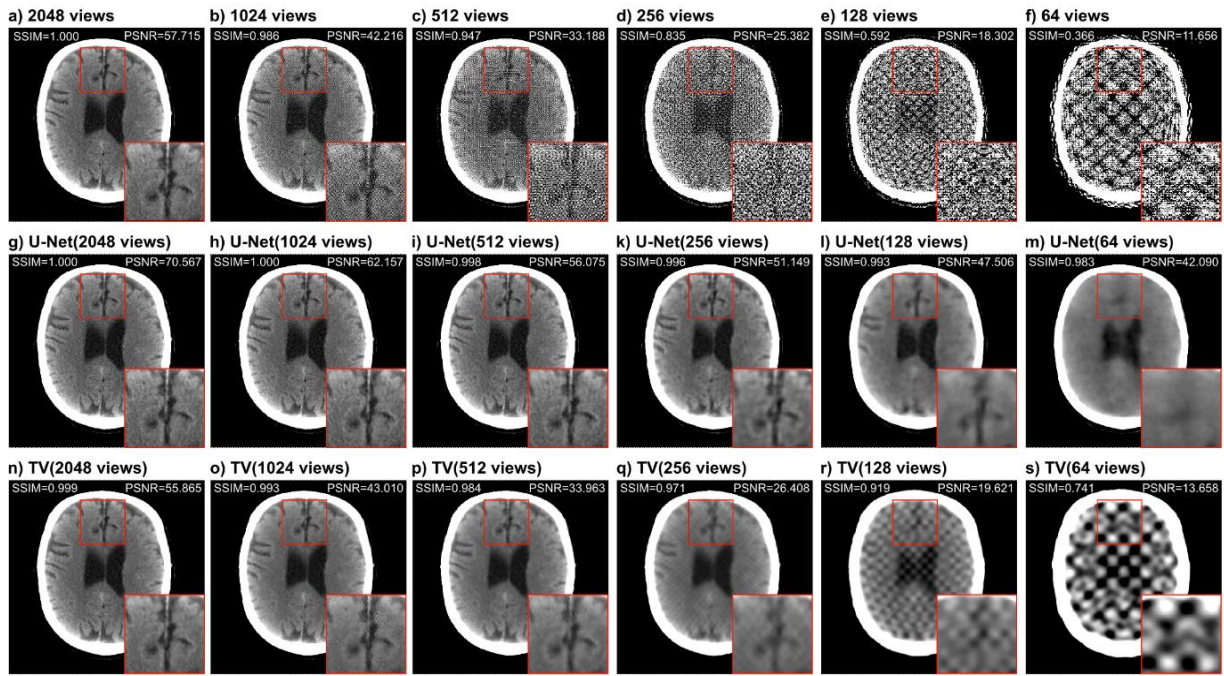

*Figure 4:* CT image (512x512 pixels) from the test set labeled as "healthy". Images a) - f) show the FBP reconstruction from 2048, 1024, 512, 256, 128, and 64 views, respectively. Images g) - m) show the U-Net predictions of the respective images in the upper row, and images n) - s) show the results of the total variation-based method. The presented structural similarity index measure (SSIM) and Peak signal-to-noise ratio (PSNR) values were calculated over the entire CT image scaled to [0- 1] from the full Hounsfield unit range [-1024-3071] HU with respect to the 4096-view reconstruction. All images are presented in the brain window ranging from 0 HU to 80 HU. The insert is 100x100 pixels, the entire image is 512x512 pixels.

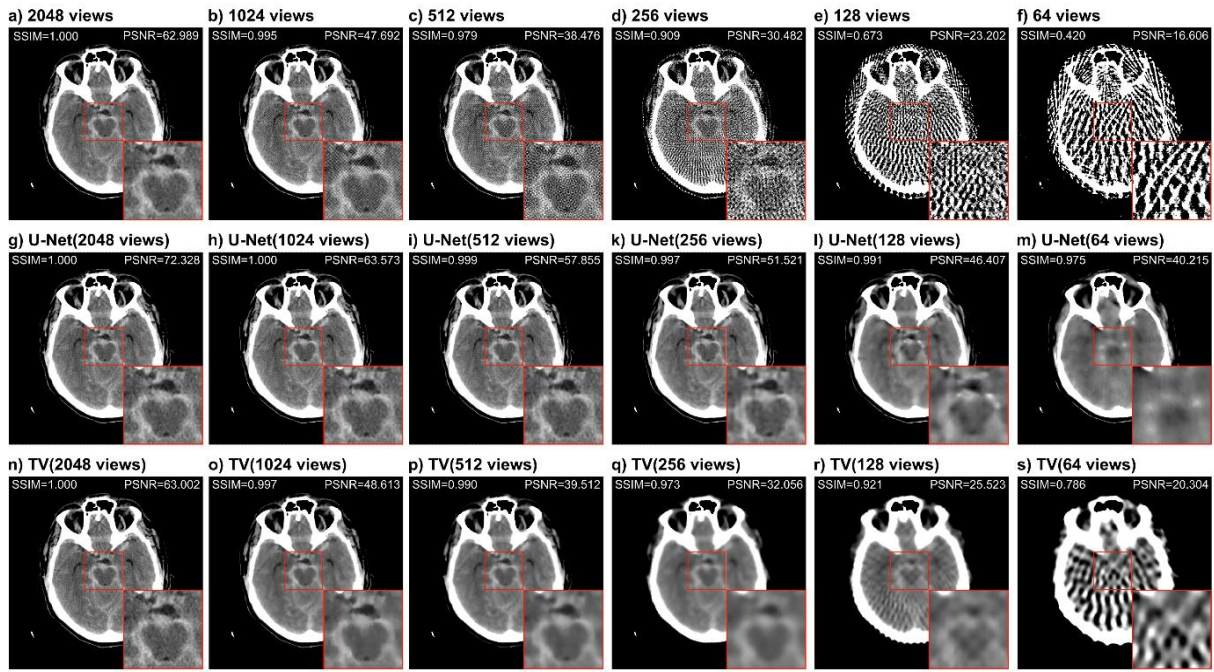

*Figure 5:* CT image (512x512 pixels) from the test set labeled with a subarachnoid hemorrhage. Images a) - f) show the FBP reconstruction from 2048, 1024, 512, 256, 128, and 64 views, respectively. Images g) - m) show the U-Net predictions of the respective images in the upper row, and images n) - s) show the results of the total variation-based method. The presented structural similarity index measure (SSIM) and Peak signal-to-noise ratio (PSNR) values were calculated over the entire CT image scaled to [0- 1] from the full Hounsfield unit range [-1024-3071] HU with respect to the 4096-view reconstruction. All images are presented in the brain window ranging from 0 HU to 80 HU. The insert is 100x100 pixels, the entire image is 512x512 pixels.

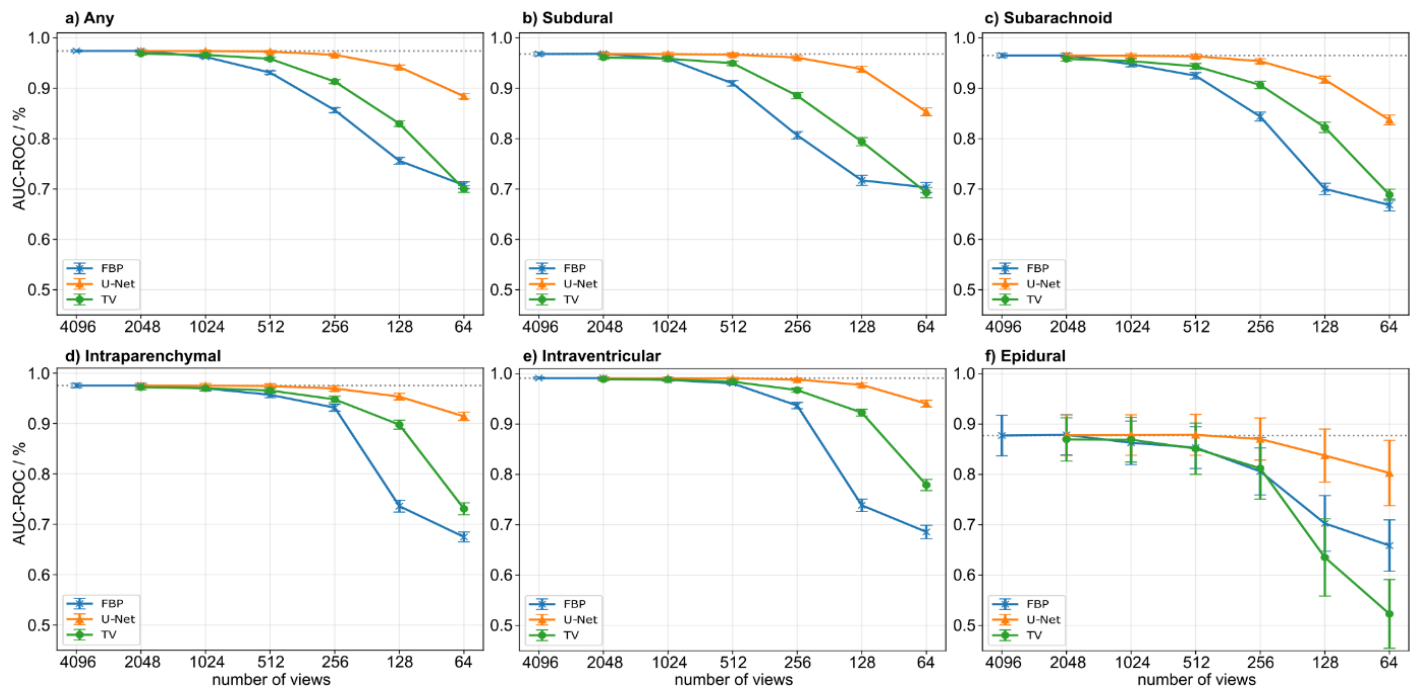

Figure 6: Results of the EfficientNetB2 detection network. Images a) - f) depict the mean area under the receiver operator characteristic curve (AUC-ROC) values (95% CI) associated with the any, subdural, subarachnoid, intraparenchymal, intraventricular, and epidural classes, respectively. The individual p-values among the different values can be found in the supplementary table S2.

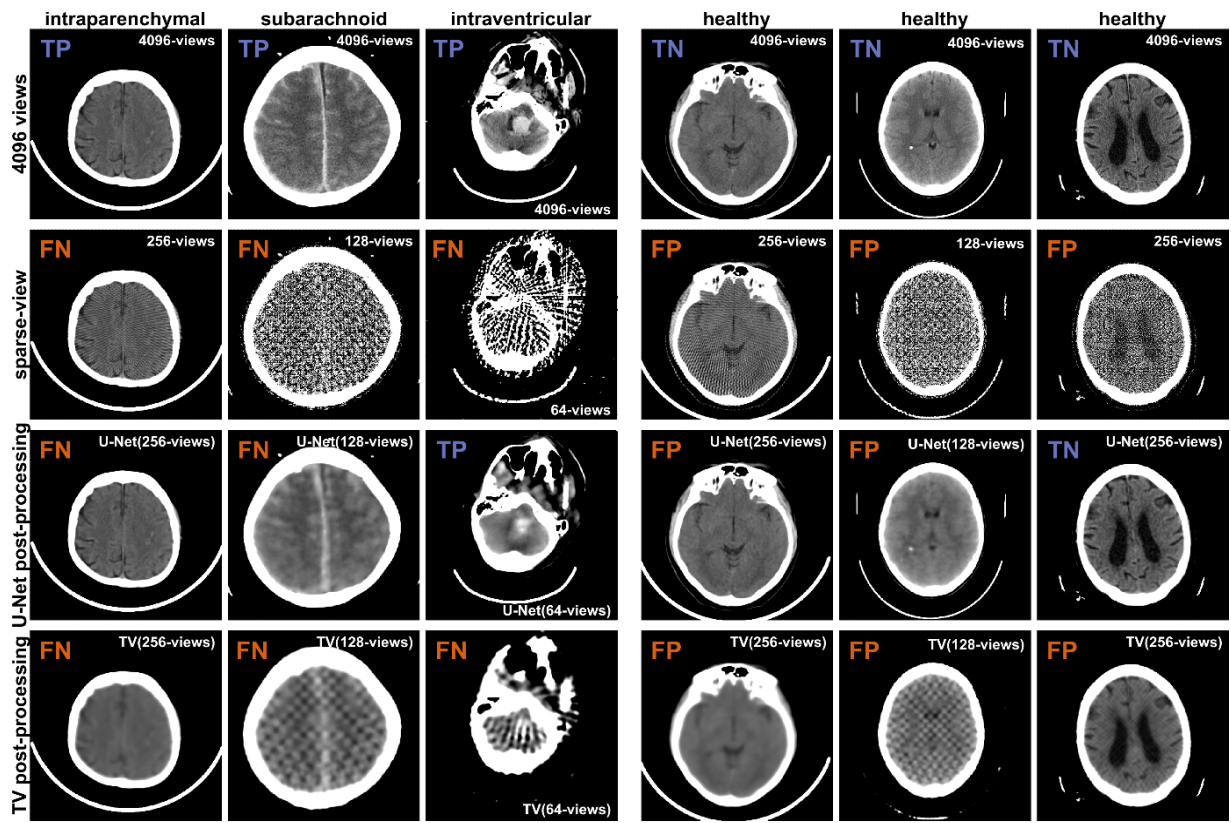

*Figure 7:* Examples of classification results of individual slices by the EfficientNetB2 detection network. Each column displays a full-view image, followed by its corresponding sparse-view reconstruction, and then the sparse-view image post-processed, either by the U-Net or the total variation (TV) algorithm. The first three columns depict images labeled with intraparenchymal, subarachnoid, and intraventricular hemorrhages, respectively, while the last three columns show healthy images. In the upper left corner of each image, it is indicated whether the images were correctly classified (true positive (TP) or true negative (TN)) or not (false negative (FN) or false positive (FP)), utilizing the same thresholds as for the confusion matrices. All images are presented in the brain window ranging from 0 HU to 80 HU.

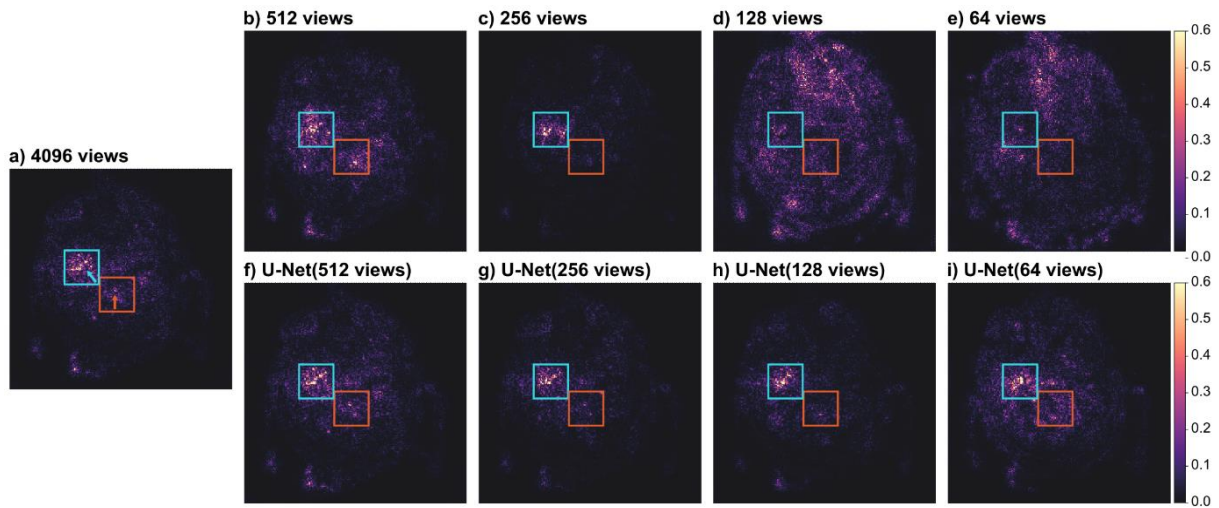

*Figure 8:* Saliency maps by the EfficientNetB2 model of the CT images of figure 3 with regards to the 'any' class. Image a) shows the saliency map of the full-view image. Images b) - e) show the saliency map of the images reconstructed from 512, 256, 128, and 64 views, respectively. Images e)- h) show the saliency maps of the images post-processed by the U-Net of the corresponding sparse-view images. All maps were normalized via min-max normalization to range [0-1]. The rectangles are at the same position as in figure 3, indicating the location of the present hemorrhages.

Explanatory note: Saliency maps were generated by analyzing the gradients of the EfficientNetB2 model with respect to the input. High values indicate that changes to those pixels have a substantial impact on the model's output, and therefore, those pixels are most important for the prediction.

# Supplementary Material

*Table S1:* Area under the receiver operator characteristics curve (AUC-ROC) values of the hemorrhage subtypes of the full-view images and images of varying levels of subsampling from 2048 to 64 views, without post-processing (FBP) and with either TV post-processing or U-Net post-processing with 95% confidence intervals in brackets.

| subtype | Nr. of views | **4096** | **2048** | **1024** | **512** | **256** | **128** | **64** |
|---|---|---|---|---|---|---|---|---|
| any | **FBP** | 0.974 (0.972-0.976) | 0.974 (0.972-0.976) | 0.962 (0.960-0.965) | 0.931 (0.928-0.935) | 0.857 (0.852-0.862) | 0.756 (0.749-0.763) | 0.708 (0.701-0.714) |
| any | **U-Net** | x | 0.974 (0.972-0.976) | **0.974 (0.972-0.976)** | **0.973 (0.971-0.975)** | **0.967 (0.964-0.969)** | **0.942 (0.939-0.946)** | **0.884 (0.879-0.889)** |
| any | **TV** | x | 0.969 (0.966-0.971) | 0.966 (0.964-0.969) | 0.958 (0.955-0.961) | 0.913 (0.909-0.917) | 0.830 (0.824-0.835) | 0.700 (0.693-0.707) |
| subdural | **FBP** | 0.968 (0.965-0.972) | 0.968 (0.965-0.972) | 0.958 (0.954-0.962) | 0.910 (0.904-0.915) | 0.807 (0.799-0.814) | 0.717 (0.707-0.727) | 0.703 (0.693-0.713) |
| subdural | **U-Net** | x | 0.968 (0.965-0.972) | **0.968 (0.964-0.971)** | **0.967 (0.963-0.970)** | **0.961 (0.957-0.965)** | **0.938 (0.933-0.943)** | **0.853 (0.845-0.861)** |
| subdural | **TV** | x | 0.961 (0.957-0.965) | 0.959 (0.955-0.963) | 0.950 (0.945-0.954) | 0.885 (0.879-0.891) | 0.794 (0.786-0.802) | 0.693 (0.683-0.703) |
| subarachnoid | FBP | 0.965 (0.961-0.969) | 0.965 (0.961-0.969) | 0.948 (0.943-0.953) | 0.925 (0.919-0.931) | 0.844 (0.835-0.853) | 0.700 (0.689-0.711) | 0.668 (0.657-0.680) |
| subarachnoid | U-Net | x | 0.965 (0.961-0.969) | **0.965 (0.961-0.969)** | **0.963 (0.959-0.968)** | **0.954 (0.949-0.959)** | **0.917 (0.910-0.924)** | **0.837 (0.827-0.847)** |
| subarachnoid | TV | x | 0.958 (0.954-0.963) | 0.954 (0.949-0.959) | 0.944 (0.938-0.949) | 0.907 (0.899-0.914) | 0.822 (0.812-0.833) | 0.689 (0.677-0.700) |
| intraparenchymal | **FBP** | 0.976 (0.971-0.980) | 0.976 (0.971-0.980) | 0.970 (0.965-0.975) | 0.957 (0.951-0.963) | 0.931 (0.925-0.938) | 0.736 (0.724-0.748) | 0.675 (0.665-0.685) |
| intraparenchymal | **U-Net** | x | 0.976 (0.971-0.980) | **0.975 (0.971-0.979)** | **0.974 (0.970-0.979)** | **0.970 (0.965-0.974)** | **0.953 (0.947-0.960)** | **0.914 (0.906-0.922)** |
| intraparenchymal | **TV** | x | 0.972 (0.967-0.977) | 0.970 (0.965-0.975) | 0.966 (0.960-0.971) | 0.948 (0.942-0.954) | 0.898 (0.889-0.907) | 0.731 (0.719-0.743) |
| intraventricular | **FBP** | 0.991 (0.989-0.993) | 0.991 (0.989-0.993) | 0.988 (0.985-0.990) | 0.981 (0.978-0.984) | 0.937 (0.930-0.943) | 0.739 (0.726-0.751) | 0.686 (0.673-0.699) |
| intraventricular | **U-Net** | x | 0.991 (0.989-0.993) | **0.991 (0.989-0.993)** | **0.991 (0.988-0.993)** | **0.988 (0.986-0.991)** | **0.978 (0.974-0.981)** | **0.941 (0.934-0.947)** |
| intraventricular | **TV** | x | 0.989 (0.987-0.991) | 0.988 (0.986-0.991) | 0.984 (0.981-0.987) | 0.967 (0.963-0.972) | 0.923 (0.916-0.929) | 0.779 (0.768-0.790) |
| epidural | **FBP** | 0.877 (0.837-0.918) | **0.879 (0.839-0.918)** | 0.863 (0.820-0.906) | 0.853 (0.812-0.895) | 0.806 (0.759-0.852) | 0.703 (0.647-0.758) | 0.659 (0.608-0.710) |
| epidural | **U-Net** | x | 0.878 (0.837-0.918) | **0.878 (0.838-0.918)** | **0.879 (0.838-0.919)** | **0.870 (0.828-0.912)** | **0.837 (0.785-0.890)** | **0.803 (0.738-0.868)** |
| epidural | **TV** | x | 0.869 (0.827-0.912) | 0.869 (0.825-0.913) | 0.851 (0.800-0.902) | 0.812 (0.751-0.873) | 0.635 (0.558-0.712) | 0.523 (0.455-0.591) |

*Table S2:* P-values of ROC-AUC values depicted in Figure 5 and Table S1 estimated using De Long's two-sided test. Statistical differences among ROC-AUC values without post-processing (Filtered backprojection (FBP)), with total variation (TV) post-processing, and U-Net post-processing were compared, both within each other and against the ROC-AUC values of the full-view reconstruction.

| subtype | Nr. of views | 2048 | 1024 | 512 | 256 | 128 | 64 |
|---|---|---|---|---|---|---|---|
| any | FBP vs. full-view | 0.02051 | <0.00017 | <0.00017 | <0.00017 | <0.00017 | <0.00017 |
| any | U-Net vs. full-view | 0.10259 | <0.00017 | <0.00017 | <0.00017 | <0.00017 | <0.00017 |
| any | TV vs. full-view | <0.00017 | <0.00017 | <0.00017 | <0.00017 | <0.00017 | <0.00017 |
| any | FBP vs. TV | <0.0003 | <0.0003 | <0.0003 | <0.0003 | <0.0003 | 0.11566 |
| any | FBP vs. U-Net | 0.00162 | <0.0003 | <0.0003 | <0.0003 | <0.0003 | <0.0003 |
| any | TV vs. U-Net | <0.0003 | <0.0003 | <0.0003 | <0.0003 | <0.0003 | <0.0003 |
| subdural | FBP vs. full-view | <0.00017 | <0.00017 | <0.00017 | <0.00017 | <0.00017 | <0.00017 |
| subdural | U-Net vs. full-view | 0.33821 | 0.12383 | <0.00017 | <0.00017 | <0.00017 | <0.00017 |
| subdural | TV vs. full-view | <0.00017 | <0.00017 | <0.00017 | <0.00017 | <0.00017 | <0.00017 |
| subdural | FBP vs. TV | <0.0003 | 0.84784 | <0.0003 | <0.0003 | <0.0003 | 0.20193 |
| subdural | FBP vs. U-Net | <0.0003 | <0.0003 | <0.0003 | <0.0003 | <0.0003 | <0.0003 |
| subdural | TV vs. U-Net | <0.0003 | <0.0003 | <0.0003 | <0.0003 | <0.0003 | <0.0003 |
| subarachnoid | FBP vs. full-view | 0.85171 | <0.00017 | <0.00017 | <0.00017 | <0.00017 | <0.00017 |
| subarachnoid | U-Net vs. full-view | 0.41305 | <0.00017 | <0.00017 | <0.00017 | <0.00017 | <0.00017 |
| subarachnoid | TV vs. full-view | <0.00017 | <0.00017 | <0.00017 | <0.00017 | <0.00017 | <0.00017 |
| subarachnoid | FBP vs. TV | <0.0003 | <0.0003 | <0.0003 | <0.0003 | <0.0003 | 0.01207 |
| subarachnoid | FBP vs. U-Net | 0.55736 | <0.0003 | <0.0003 | <0.0003 | <0.0003 | <0.0003 |
| subarachnoid | TV vs. U-Net | <0.0003 | <0.0003 | <0.0003 | <0.0003 | <0.0003 | <0.0003 |
| intraparenchymal | FBP vs. full-view | 0.2601 | <0.00017 | <0.00017 | <0.00017 | <0.00017 | <0.00017 |
| intraparenchymal | U-Net vs. full-view | 0.28184 | <0.00017 | <0.00017 | <0.00017 | <0.00017 | <0.00017 |
| intraparenchymal | TV vs. full-view | <0.00017 | <0.00017 | <0.00017 | <0.00017 | <0.00017 | <0.00017 |
| intraparenchymal | FBP vs. TV | <0.0003 | 0.63932 | <0.0003 | <0.0003 | <0.0003 | <0.0003 |
| intraparenchymal | FBP vs. U-Net | 0.14081 | <0.0003 | <0.0003 | <0.0003 | <0.0003 | <0.0003 |
| intraparenchymal | TV vs. U-Net | <0.0003 | <0.0003 | <0.0003 | <0.0003 | <0.0003 | <0.0003 |
| intraventricular | FBP vs. full-view | <0.00017 | <0.00017 | <0.00017 | <0.00017 | <0.00017 | <0.00017 |
| intraventricular | U-Net vs. full-view | 0.9009 | <0.00017 | <0.00017 | <0.00017 | <0.00017 | <0.00017 |
| intraventricular | TV vs. full-view | <0.00017 | <0.00017 | <0.00017 | <0.00017 | <0.00017 | <0.00017 |
| intraventricular | FBP vs. TV | <0.0003 | 0.50299 | 0.00881 | <0.0003 | <0.0003 | <0.0003 |
| intraventricular | FBP vs. U-Net | 0.00032 | <0.0003 | <0.0003 | <0.0003 | <0.0003 | <0.0003 |
| intraventricular | TV vs. U-Net | <0.0003 | <0.0003 | <0.0003 | <0.0003 | <0.0003 | <0.0003 |
| epidural | FBP vs. full-view | 0.01858 | 0.01238 | <0.00017 | <0.00017 | <0.00017 | <0.00017 |
| epidural | U-Net vs. full-view | 0.20895 | 0.45048 | 0.61676 | 0.17415 | <0.00017 | <0.00017 |
| epidural | TV vs. full-view | 0.10562 | 0.01618 | 0.00032 | <0.00017 | <0.00017 | <0.00017 |
| epidural | FBP vs. TV | 0.07121 | 0.45476 | 0.7749 | 0.66772 | 0.01521 | <0.0003 |
| epidural | FBP vs. U-Net | 0.17609 | 0.02339 | <0.0003 | <0.0003 | <0.0003 | <0.0003 |
| epidural | TV vs. U-Net | 0.07476 | 0.00101 | <0.0003 | <0.0003 | <0.0003 | <0.0003 |

Table S3: Confusion matrices of the hemorrhage subtypes of the full-view images and images of varying levels of subsampling from 2048 to 64 views, without post-processing (FBP) and with either TV post-processing or U-Net post-processing.

| Nr. of views | | | 4096 | | 2048 | | 1024 | | 512 | | 256 | | 128 | | 64 | |
|---|---|---|---|---|---|---|---|---|---|---|---|---|---|---|---|---|
| subtype | type | actual / predicted | yes | no | yes | no | yes | no | yes | no | yes | no | yes | no | yes | no |
| any | FBP | yes | 4684 | 2287 | 4703 | 2456 | 4588 | 3343 | 4431 | 6580 | 4167 | 9398 | 3720 | 11937 | 3630 | 12952 |
| | | no | 455 | 31999 | 436 | 31830 | 551 | 30943 | 708 | 27706 | 972 | 24888 | 1419 | 22349 | 1509 | 21334 |
| | U-Net | yes | | | 4686 | 2293 | 4659 | 2104 | 4673 | 2388 | 4568 | 2541 | 4344 | 3277 | 4025 | 6287 |
| | | no | | | 453 | 31993 | 480 | 32182 | 466 | 31898 | 571 | 31745 | 795 | 31009 | 1114 | 27999 |
| | TV | yes | | | 4607 | 2370 | 4585 | 2390 | 4486 | 2697 | 4309 | 6129 | 3887 | 9152 | 3691 | 13881 |
| | | no | | | 532 | 31916 | 554 | 31896 | 653 | 31589 | 830 | 28157 | 1252 | 25134 | 1448 | 20405 |
| subdural | FBP | yes | 1994 | 3185 | 2003 | 3369 | 1998 | 4364 | 1933 | 8634 | 1912 | 14489 | 1489 | 13628 | 1448 | 13502 |
| | | no | 209 | 34037 | 200 | 33853 | 205 | 32858 | 270 | 28588 | 291 | 22733 | 714 | 23594 | 755 | 23720 |
| | U-Net | yes | | | 1984 | 2988 | 1997 | 3252 | 1990 | 3294 | 1949 | 3453 | 1884 | 4632 | 1704 | 8466 |
| | | no | | | 219 | 34234 | 206 | 33970 | 213 | 33928 | 254 | 33769 | 319 | 32590 | 499 | 28756 |
| | TV | yes | | | 1925 | 2848 | 1906 | 2926 | 1937 | 4339 | 1872 | 8718 | 1781 | 13284 | 1449 | 13604 |
| | | no | | | 278 | 34374 | 297 | 34296 | 266 | 32883 | 331 | 28504 | 422 | 23938 | 754 | 23618 |
| subarachnoid | FBP | yes | 1578 | 2612 | 1588 | 2888 | 1537 | 3995 | 1405 | 3885 | 1324 | 8761 | 1159 | 13452 | 1094 | 14489 |
| | | no | 204 | 35031 | 194 | 34755 | 245 | 33648 | 377 | 33758 | 458 | 28882 | 623 | 24191 | 688 | 23154 |
| | U-Net | yes | | | 1588 | 2812 | 1586 | 2809 | 1583 | 2848 | 1537 | 3182 | 1464 | 4857 | 1338 | 8247 |
| | | no | | | 194 | 34831 | 196 | 34834 | 199 | 34795 | 245 | 34461 | 318 | 32786 | 444 | 29396 |
| | TV | yes | | | 1549 | 2812 | 1532 | 2973 | 1533 | 4043 | 1431 | 5357 | 1288 | 9354 | 1154 | 14011 |
| | | no | | | 233 | 34831 | 250 | 34670 | 249 | 33600 | 351 | 32286 | 494 | 28289 | 628 | 23632 |
| intraparenchymal | FBP | yes | 1637 | 1527 | 1636 | 1529 | 1622 | 1770 | 1542 | 2004 | 1460 | 3659 | 1171 | 12579 | 1272 | 16257 |
| | | no | 149 | 36112 | 150 | 36110 | 164 | 35869 | 244 | 35635 | 326 | 33980 | 615 | 25060 | 514 | 21382 |
| | U-Net | yes | | | 1639 | 1551 | 1636 | 1553 | 1619 | 1207 | 1606 | 1428 | 1589 | 2666 | 1499 | 5471 |
| | | no | | | 147 | 36088 | 150 | 36086 | 167 | 36432 | 180 | 36211 | 197 | 34973 | 287 | 32168 |
| | TV | yes | | | 1609 | 1213 | 1616 | 1423 | 1611 | 1921 | 1542 | 2384 | 1387 | 3790 | 1183 | 12355 |
| | | no | | | 177 | 36426 | 170 | 36216 | 175 | 35718 | 244 | 35255 | 399 | 33849 | 603 | 25284 |
| intraventricular | FBP | yes | 1143 | 1244 | 1145 | 1316 | 1135 | 1815 | 1121 | 2978 | 1044 | 6526 | 817 | 14000 | 791 | 15983 |
| | | no | 59 | 36979 | 57 | 36907 | 67 | 36408 | 81 | 35245 | 158 | 31697 | 385 | 24223 | 411 | 22240 |
| | U-Net | yes | | | 1145 | 1335 | 1144 | 1328 | 1145 | 1356 | 1123 | 1195 | 1103 | 2475 | 1055 | 5061 |
| | | no | | | 57 | 36888 | 58 | 36895 | 57 | 36867 | 79 | 37028 | 99 | 35748 | 147 | 33162 |
| | TV | yes | | | 1130 | 1158 | 1132 | 1447 | 1130 | 2124 | 1077 | 3059 | 1026 | 6534 | 889 | 11824 |
| | | no | | | 72 | 37065 | 70 | 36776 | 72 | 36099 | 125 | 35164 | 176 | 31689 | 313 | 26399 |
| epidural | FBP | yes | 66 | 2148 | 66 | 2149 | 66 | 2703 | 63 | 4336 | 66 | 9040 | 61 | 12805 | 56 | 14648 |
| | | no | 25 | 37186 | 25 | 37185 | 25 | 36631 | 28 | 34998 | 25 | 30294 | 30 | 26529 | 35 | 24686 |
| | U-Net | yes | | | 66 | 2134 | 66 | 2161 | 66 | 2359 | 66 | 2831 | 65 | 2521 | 65 | 4378 |
| | | no | | | 25 | 37200 | 25 | 37173 | 25 | 36975 | 25 | 36503 | 26 | 36813 | 26 | 34956 |
| | TV | yes | | | 66 | 1912 | 66 | 1815 | 66 | 2173 | 64 | 3245 | 41 | 3255 | 42 | 12137 |
| | | no | | | 25 | 37422 | 25 | 37519 | 25 | 37161 | 27 | 36089 | 50 | 36079 | 49 | 27197 |

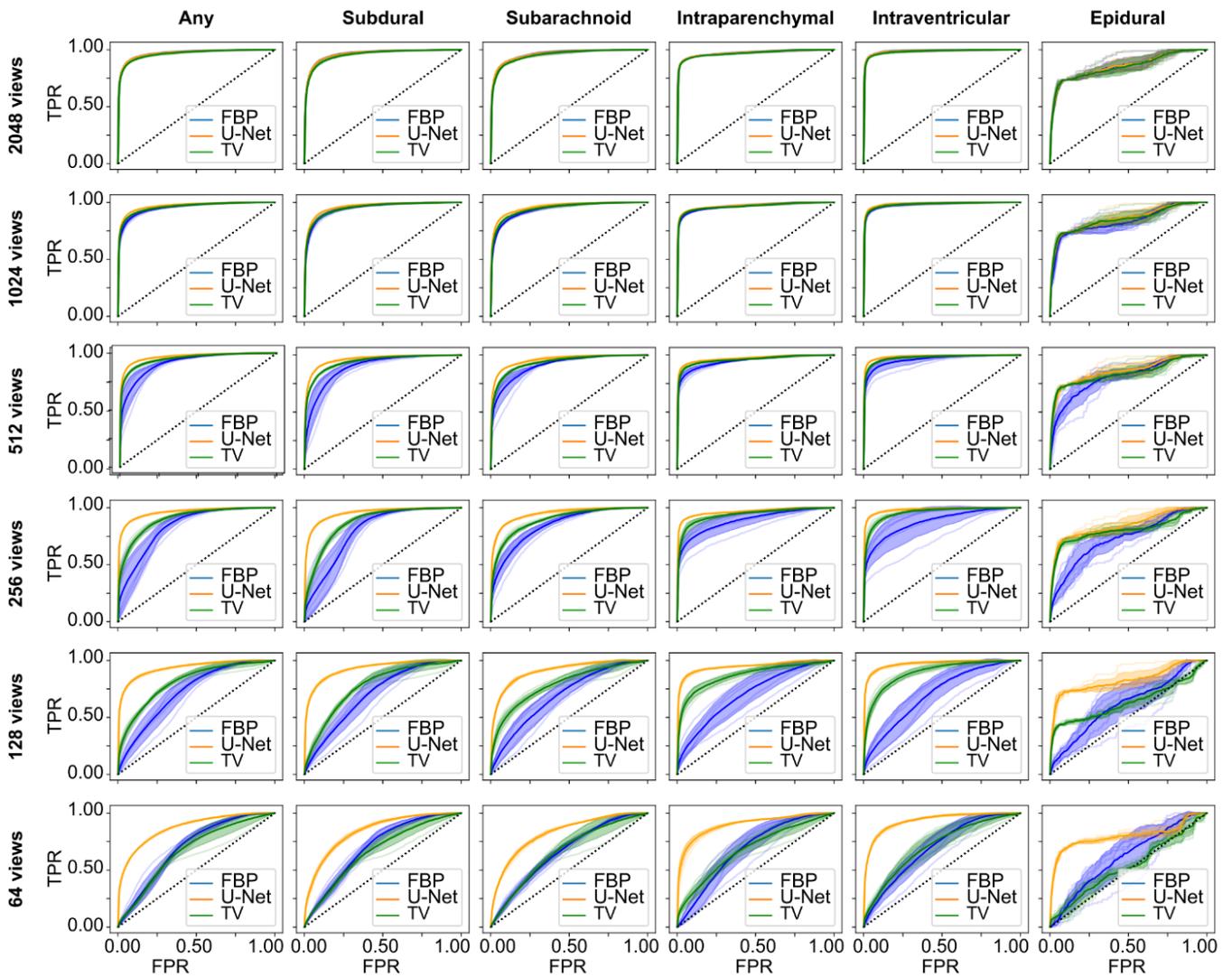

*Figure S1:* Results of the EfficientNetB2 hemorrhage classification network. Depicted are the mean receiver operator characteristic curves of the classes any, subdural, subarachnoid, intraparenchymal, intraventricular, and epidural, for the different levels of subsampling, which were used for the calculation of the individual AUC values. In addition to the results of the raw images (blue), the classification results on the images post-processed by either U-Net (orange) or the total variation-based method (green) are shown. The bold graphs are the mean curves, the colored areas mark the 95% confidence intervals.

| SNR [dB] | 2048 views | 1024 views | 512 views | 256 views | 128 views | 64 views |
|---|---|---|---|---|---|---|
| FBP | 48.02 | 36.80 | 31.05 | 20.99 | 12.39 | 6.25 |
| U-Net | **51.82** | **46.24** | **42.39** | **36.83** | **32.49** | **28.36** |
| TV | 40.24 | 36.90 | 33.78 | 26.01 | 18.57 | 12.05 |

*Table S4*: Signal-to-noise ratios (SNRs) of the RSNA test set. The images were reconstructed by filtered backprojection (FBP) from varying number of views and then not post-processing, post-processed by the U-Net, or post-processed by the total variation-based method (TV).